# Machine Learning and Deep Learning in Quantum Materials: Symmetry, Topology, and the Rise of Altermagnets

**Mahyar Hassani-Vasmejani[1], Hosein Alavi-Rad[2*], Meysam Bagheri Tagani[3]**
[1]Department of Computer Engineering, Lan.C., Islamic Azad University, Langarud, Iran
mahyar.hassanivasmejani@iau.ir
[2]Department of Electrical Engineering, Lan.C., Islamic Azad University, Langarud, Iran
ho.alavirad@iau.ac.ir
[3]Department of Physics, University of Guilan, P.O. Box 41335-1914, Rasht, Iran
m_bagheri@guilan.ac.ir

**Abstract**

The contemporary landscape of condensed matter physics is currently navigating a data deluge of unprecedented scale, driven by the proliferation of high-throughput ab initio calculations and the exponential growth of experimental datasets from large-scale facilities. While traditional first-principles methods like Density Functional Theory (DFT) have served as the bedrock of materials science for decades, their intrinsic cubic scaling with system size ($O(N^3)$) creates an insurmountable bottleneck when attempting to scan the vast, combinatorial chemical space of potential quantum materials. This review provides an exhaustive analysis of how Machine Learning (ML) and Deep Learning (DL) are transcending these computational limitations to accelerate the discovery of exotic phases of matter. We critically examine the transition from rigid, descriptor-based approaches to flexible, symmetry-aware architectures, such as E(3)-equivariant Graph Neural Networks (GNNs), which respect the fundamental physical laws of rotation and translation. A central focus is placed on the automated discovery of topological phases, where ML models leverage symmetry indicators and elementary band representations to diagnose non-trivial topology without the need for expensive band structure integration. The narrative culminates in a detailed case study of the Altermagnet, a newly identified third branch of magnetism that defies the conventional ferromagnetic-antiferromagnetic dichotomy. We detail how specialized AI search engines, integrating graph theory with crystallographic symmetry analysis, have successfully identified d-wave, g-wave, and the elusive i-wave altermagnets, fundamentally expanding our understanding of magnetic order. The report concludes by addressing the critical interpretability gap, advocating for symbolic regression and active learning workflows to bridge the divide between black-box predictions and experimental verification.

# 1. Introduction

## 1.1 The Quantum Many-Body Problem and the Data Bottleneck

The central intellectual challenge in modern condensed matter physics is the quantum many-body problem: predicting the collective behavior of interacting electrons within a crystal lattice. This problem is defined by a level of complexity that scales exponentially with the number of particles, making exact solutions impossible for macroscopic systems. For the past half-century, the Kohn-Sham formulation of Density Functional Theory (DFT) has provided a pragmatic and remarkably successful solution, reducing the interacting many-body problem to a system of non-interacting particles moving in an effective mean field. DFT has become the standard model of materials science, allowing researchers to predict ground-state energies, lattice constants, and electronic band structures with high accuracy [1-3].

However, the success of DFT has birthed a new crisis. The computational cost of standard DFT calculations scales roughly as the cube of the number of electrons ($O(N^3)$), rendering the simulation of large supercells, disordered systems, or complex magnetic textures prohibitively expensive [4]. While ML offers a surrogate, high-fidelity ground truth data remains essential; recent advances [5] demonstrate that many-body Quantum Monte Carlo (QMC) methods can surpass standard DFT in property prediction for 2D materials, providing a higher-accuracy tier of training data for future architectures. When one considers the combinatorial vastness of chemical space, estimated to contain upwards of $10^{60}$ possible distinct pharmacological molecules and an equally daunting number of stable inorganic crystal structures, it becomes evident that traditional methods are insufficient for comprehensive exploration [6, 7]. The search for materials with specific, exotic functionalities, such as high-temperature superconductivity or non-trivial topology, requires scanning this space with a speed and efficiency that DFT simply cannot provide.

Simultaneously, the scientific community faces a data deluge of petabyte-scale proportions. Large-scale experimental facilities, such as the Large Hadron Collider (LHC) and the Square Kilometre Array (SKA), generate data rates that exceed global internet traffic, requiring automated pipelines for real-time analysis [8, 9]. In the realm of materials science, this is paralleled by the creation of massive computational databases like the Materials Project, AFLOW, and the Topological Materials Database. These repositories contain electronic structure data for hundreds of thousands of compounds, yet they represent only a fraction of the chemically feasible space. The sheer volume of data has outpaced human intuition, necessitating algorithmic approaches to identify hidden patterns and correlations that are invisible to the naked eye [10].

This bottleneck has spurred the widespread adoption of Machine Learning (ML) as a surrogate for quantum mechanics. By learning the non-linear mapping from atomic structure to physical property (structure-property relationships) or from potential to energy (potential energy surfaces), ML models can predict material properties orders of magnitude faster than DFT [11, 12]. However, the application of ML to quantum materials is distinct from standard computer vision or natural language processing tasks; it requires a rigorous adherence to physical symmetries and constraints [13, 14]. The field is currently witnessing a turf battle and a convergence between classical AI and quantum computing, where AI models trained on classical data are beginning to outperform quantum simulations for weakly correlated systems, while quantum algorithms promise eventual supremacy for strongly correlated regimes [15, 16].

## 1.2 The Symmetry-Topology Connection

A transformative development in condensed matter physics over the last two decades has been the classification of materials not just by their symmetry breaking (the Landau paradigm) but by their quantum topology [17, 18]. Topological phases of matter, such as Topological Insulators (TIs), Weyl Semimetals (WSMs), and Axion Insulators, are defined by global invariants, such as Chern numbers or $\mathbb{Z}_2$ indices, that remain robust against local perturbations and disorder [19]. These phases host exotic boundary states, such as the conducting surface states of a TI or the Fermi arcs of a WSM, which are protected by the bulk topology.

The connection between crystal symmetry and topology is profound and has become the linchpin of modern materials discovery. The theory of Topological Quantum Chemistry (TQC) and Symmetry-Based Indicators (SIs) established that the topological character of a band structure is largely determined by the symmetry representations of the occupied bands at high-symmetry points in the Brillouin zone [20, 21]. This realization converted the search for topological materials from a problem requiring full Brillouin zone integration (calculation of Berry curvature) to a discrete classification problem based on space group data [22]. As summarized in Figure 1, this reduction in complexity makes topological phase discovery an ideal candidate for ML approaches, which can rapidly process crystallographic data to predict topological invariants without running expensive wave-function calculations [23].

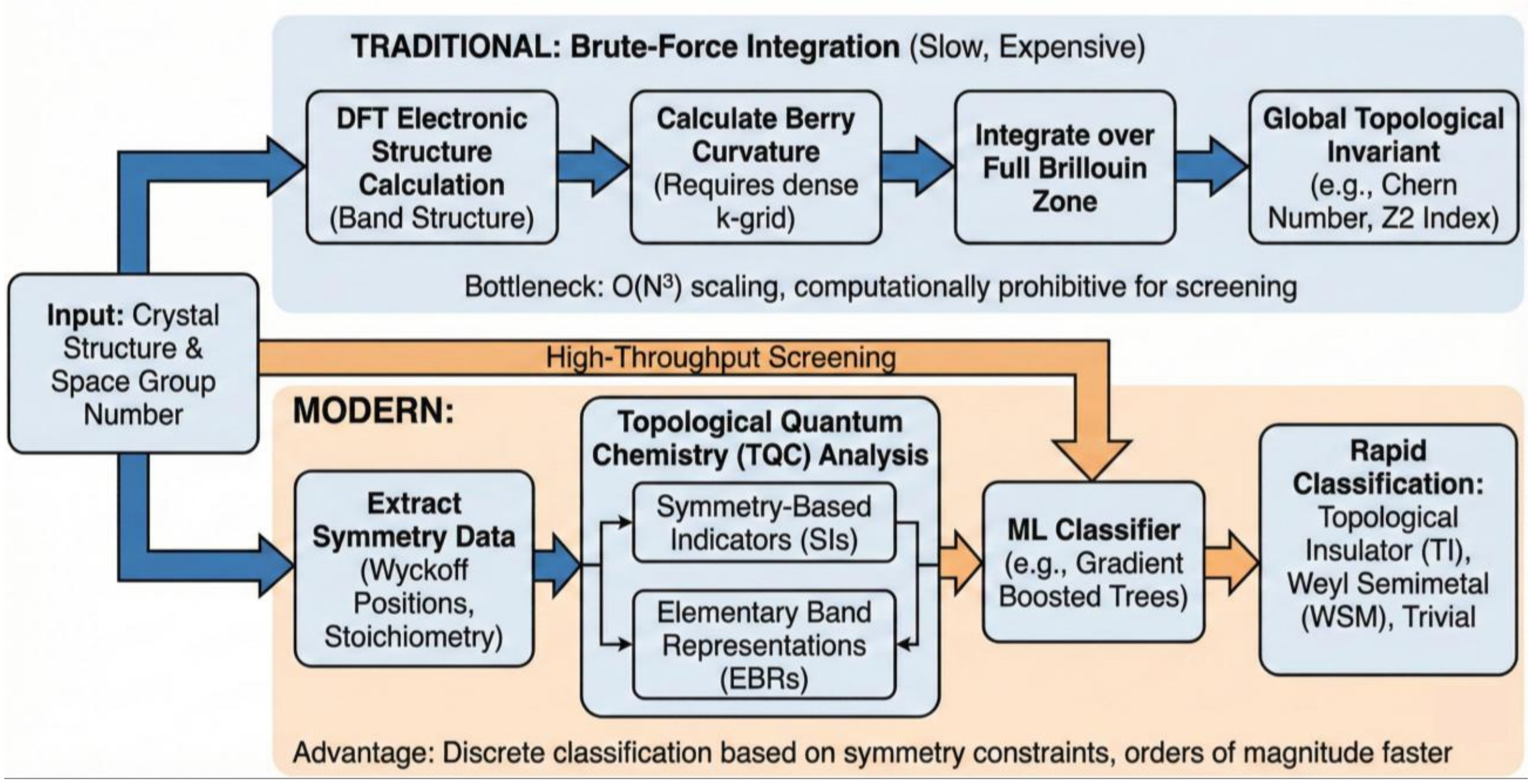

Figure 1. High-Throughput Discovery of Topological Phases.

## 1.3 The Rise of Altermagnets

While ferromagnetism and antiferromagnetism have been known for centuries and nearly a century, respectively, a third fundamental branch of magnetism has recently emerged: Altermagnetism. This discovery fundamentally disrupts the conventional binary classification of magnetic order. Altermagnets defy the traditional dichotomy: like antiferromagnets (AFMs), they possess zero net macroscopic magnetization; however, unlike AFMs, they exhibit large, spin-split Fermi surfaces characteristic of ferromagnets (FMs) [24].

This unique phenomenology arises from a specific interplay of crystal symmetries, typically involving rotation symmetries combined with time-reversal translation, that enforce anisotropic spin splitting in momentum space [25]. This splitting can take the form of d-wave, g-wave, or even i-wave patterns, analogous to the unconventional order parameters found in high-temperature superconductors [26]. The discovery of altermagnets was not accidental but the result of rigorous symmetry analysis and group theory. However, identifying candidate materials within the vast database of known magnetic structures requires the high-dimensional pattern-matching capabilities of deep learning. ML models, specifically designed to be sensitive to the subtle symmetry operations defining spin groups, are now driving the exploration of this new magnetic frontier [27].

## 1.4 Scope of this Review

This report provides a comprehensive examination of the intersection between machine learning and quantum materials, with a specific focus on the role of symmetry and topology. Section 2 details the evolution of neural network architectures in physics, emphasizing the shift from invariant descriptors to equivariant Graph Neural Networks (GNNs) that know physics. Section 3 explores the application of these architectures to the automated discovery of topological phases, highlighting the success of symmetry-based machine learning in classifying thousands of materials. Section 4 presents a deep dive into altermagnets, discussing the physics of anisotropic spin splitting and the specific ML workflows used to discover new candidates like MnTe and $RuO_2$. Finally, Section 5 addresses the critical challenges of interpretability and data scarcity, proposing active learning and symbolic regression as necessary paths forward.

# 2. ML/DL in Condensed Matter Physics

The success of any machine learning model in the physical sciences depends critically on how the physical system is represented. Unlike images, which are defined on a regular pixel grid, molecules and crystals are defined by continuous atomic coordinates and discrete chemical species, subject to fundamental permutation, rotation, and translation symmetries. A model that predicts different energies for a molecule when it is merely rotated is physically invalid [28].

## 2.1 Data Representation and Feature Engineering

Early approaches to materials ML relied heavily on feature engineering, where domain expertise was used to construct fixed-length vectors (descriptors) representing the material. These descriptors aimed to encode the relevant physics while satisfying invariance requirements [29].

The Coulomb Matrix is one of the earliest global descriptors, designed to encode the electrostatic interaction between atoms. For a molecule with atoms i and j, the matrix elements are defined as:

$$C_{ij} = Z_i Z_j / \left| R_i - R_j \right| \quad (1)$$

Where Z is the nuclear charge, and R is the position vector. Diagonal elements typically encode the atomic potential energy, $C_{ii} = 0.5 Z_i^{2.4}$. While conceptually simple, the Coulomb matrix suffers from a lack of invariance to atom indexing (permutation). If the rows and columns are permuted, the matrix changes, but the molecule remains the same. To address this, the sorted eigenspectrum of the matrix is often used as the input feature vector, but this compression results in a loss of geometric information [30].

For periodic systems like crystals, the Sine Matrix extends this concept. It interacts atoms not just with their neighbors in the unit cell, but with their periodic images, replacing the Euclidean distance with a sine function of the coordinates relative to the lattice vectors [31]. This captures the periodicity inherent in crystalline solids but still struggles with the curse of dimensionality as system sizes grow.

To capture the nuances of chemical bonding, specifically for d- and f-block elements crucial in magnetism, the Orbital Field Matrix (OFM) was developed. The OFM represents the material based on the distribution of valence shell electrons, utilizing the Voronoi polyhedron to define the local environment [32, 33]. By encoding the coordination number and the specific atomic species of the neighbors into a matrix representation, the OFM has proven highly effective for predicting local magnetic moments and formation energies in lanthanide-transition metal alloys. It outperforms simple geometric descriptors by explicitly accounting for the angular distribution of neighbors, which dictates the crystal field splitting responsible for magnetic anisotropy [32].

While descriptors like the Coulomb matrix, OFM, and the Smooth Overlap of Atomic Positions (SOAP) are useful for scalar predictions (e.g., formation energy, band gap), they often discard critical geometric information required for predicting vector or tensor quantities (e.g., forces, spin textures, polarizability) [34]. Furthermore, they decouple the representation from the learning process. In a deep learning paradigm, it is often preferable for the model to learn its own features from the raw atomic graph, allowing it to identify non-intuitive but physically relevant correlations that human-engineered descriptors might miss [35].

## 2.2 Architectures for Quantum Materials (CNNs, GNNs, Equivariant GNNs)

The field has largely moved toward Deep Learning architectures that learn representations directly from the atomic structure graphs, evolving from image-based analogies to rigorous geometric deep learning [36].

Standard CNNs, the workhorses of computer vision, are ill-suited for the irregular, non-Euclidean geometry of crystals [37]. A crystal lattice is not a grid of pixels. However, for specific tasks like classifying quantum phases from wavefunction snapshots or Density Matrix Renormalization Group (DMRG) outputs, mapping techniques can be used [38]. Qubism is one such technique that maps quantum states or reduced density matrices into 2D images, allowing standard CNN architectures (like LeNet-5 or ResNet) to classify phases such as the Quantum Ising or Heisenberg models [39]. While effective for model Hamiltonians on regular lattices, this approach scales poorly to complex, continuous crystalline materials where atoms can be at arbitrary positions [40].

Graph Neural Networks (GNNs) represent materials naturally as graphs where atoms are nodes and chemical bonds are edges. Information is propagated through the graph via message passing operations.

- Nodes: Embed atomic features (atomic number, electronegativity, group number).
- Edges: Embed interatomic distances and bond angles.
- Update Rule: Nodes update their internal state vector by aggregating messages from their neighbors. This allows local chemical information to propagate globally across the molecule or crystal after several layers of message passing [41, 42].

Standard GNNs (like SchNet or CGCNN) are invariant to rotation. This means if you rotate the input molecule, the output scalar (energy) remains unchanged. This is correct for energy, but insufficient for vector properties [43].

A critical advancement in the last few years is the development of E(3)-equivariant GNNs [44-46]. For predicting forces (vectors) or magnetic spins (pseudovectors), the output must rotate concomitantly with the input (Figure 2). This property is called equivariance. Architectures like NequIP, Allegro, MACE, MagNet, and ChargE3Net explicitly build rotation matrices (Wigner-D matrices) into the network layers [47].

1. ChargE3Net: This architecture focuses on predicting the electron density distribution, a scalar field that must respect the underlying symmetry of the lattice. It enables the learning of higher-order equivariant features, allowing for high-fidelity predictions of electronic structure with relatively small training datasets [47].
2. MagNet and SpinGNN: These models extend equivariance to magnetic materials. MagNet learns magnetic force vectors by mapping combined atomic and spin configurations to forces computed via DFT. It embeds E(3)-equivariance to ensure that spin configurations transform consistently with atomic rotations, which is essential for correctly modeling spin-lattice coupling and magnon dynamics [48]. SpinGNN introduces specialized architectures like the Heisenberg Edge GNN (HEGNN) to model spin-spin interactions explicitly [49].

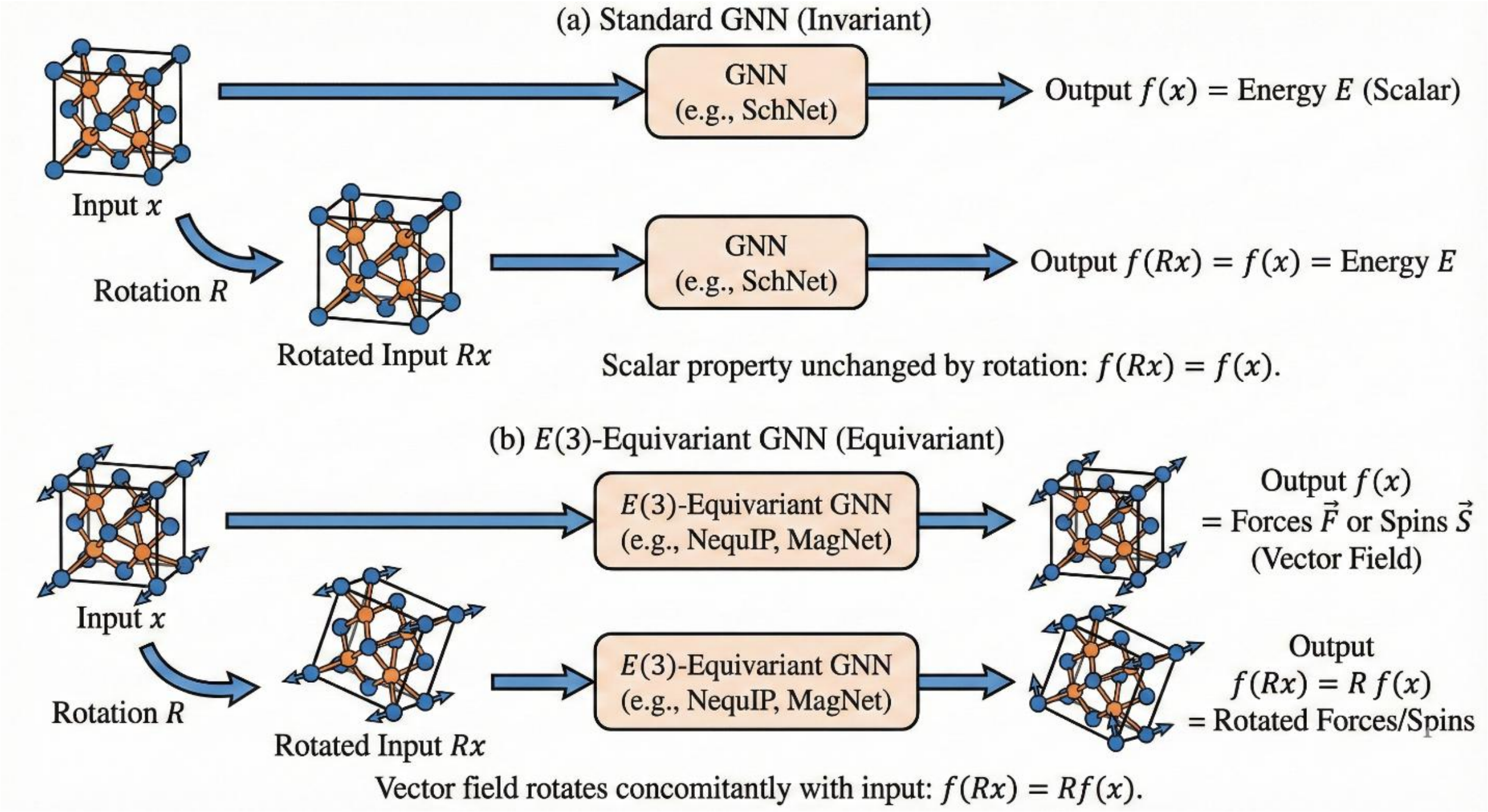

Figure 2. From Invariance to Equivariance in Graph Neural Networks. (a) Standard GNNs (e.g., SchNet) are rotationally invariant; rotating the input crystal structure leaves the predicted scalar property (e.g., Energy, E) unchanged ($f(Rx) = f(x)$). This is suitable for potential energy surfaces. (b) E(3)-Equivariant GNNs (e.g., NequIP, MagNet) explicitly encode rotation matrices into the message passing. Rotating the input results in a concomitant rotation of the output vector field (e.g., Forces, F, or Magnetic Spins, S), such that $f(Rx) = Rf(x)$. This geometric fidelity is crucial for correctly modeling

spin-lattice coupling and magnetic textures.

## 2.3 Hamiltonian Learning and Band Structure Prediction

Beyond predicting scalar properties, deep learning is now being used to predict the Hamiltonian matrix itself, the fundamental operator governing the quantum system [50].

Models like DeepH predict the tight-binding Hamiltonian elements $H_{ij}$ based on the local environment of atoms i and j. These approaches have demonstrated high accuracy in predicting Hamiltonian blocks, which is essential for capturing electronic properties, including Spin-Orbit Coupling (SOC) effects [51]. By predicting the Hamiltonian, these models allow for the derivation of all electronic properties (band structure, Density of States, Berry curvature, topological invariants) at a fraction of the computational cost of DFT [52]. The model essentially learns the rules of constructing the Hamiltonian from the atomic geometry, bypassing the iterative self-consistent field (SCF) cycles of DFT.

Other approaches map structure directly to the Density of States (DOS). For example, approaches using discretized Hamiltonians or Principal Component Analysis (PCA) on the DOS vector allow for the rapid screening of materials with desired band gaps or van Hove singularities [53-55]. These models are particularly useful for screening for specific electronic features, such as the flat bands required for heavy fermion behavior or superconductivity.

**Table 1. Comparison of Neural Network Architectures in Quantum Materials**

| **Architecture** | **Input Representation** | **Symmetry Handling** | **Best Application** | **Key Example** |
|---|---|---|---|---|
| **Standard CNN** | 2D Image / Grid (Qubism) | Translational (pixel grid) | Phase classification in model Hamiltonians | Qubism |
| **GNN (SchNet, CGCNN)** | Graph (Nodes/Edges) | Permutation Invariant | Scalar properties (Energy, Band Gap) | Materials Project API (CGCNN) |
| **E(3)-GNN (NequIP, MagNet)** | Geometric Graph | Rotation/Translation Equivariant | Vector/Tensor fields (Forces, Spins, Electron Density) | MagNet , ChargE3Net |
| **Deep** | Local Atomic | Gauge Equivariant | Electronic Band | DeepH |

| | | |
|---|---|---|
| **Hamiltonian NN** | Environment | Structure, Hamiltonian Matrix |

# 3. Topological Phase Discovery with ML

Topological materials are defined by global geometric properties of their wavefunctions, making them robust to disorder but difficult to detect with local descriptors. However, the connection between symmetry and topology has enabled a massive acceleration in their discovery via ML [56].

## 3.1 Learning Topological Invariants

The direct prediction of topological invariants (integers like the Chern number C or the $\mathbb{Z}_2$ index $\upsilon$ from the Hamiltonian is a classification task.

- Supervised Learning: Neural networks have been trained on millions of random Hamiltonians to compute the winding number (for 1D AIII class) and Chern number (for 2D A class). Surprisingly, the intermediate layers of these networks have been shown to learn quantities resembling the local Berry curvature, effectively rediscovering the mathematical formula for the invariant through backpropagation [57, 58].
- Unsupervised Learning: Techniques like k-means clustering and Variational Autoencoders (VAEs) have been applied to time-resolved ARPES data and single-particle form factors to distinguish between trivial and topological phases without labeled training data. For instance, VAEs can distinguish Fractional Chern Insulators (FCIs) from Charge Density Waves (CDWs) by analyzing the distribution of Berry curvature in the Brillouin zone. The latent space of the VAE captures the topological connectedness of the phase space, allowing for the generation of new topological Hamiltonians via interpolation [59].

## 3.2 Symmetry Indicators and TQC

The most prolific discovery workflows utilize Symmetry-Based Indicators (SIs). The theory of Topological Quantum Chemistry (TQC) posits that the topology of a material is constrained by its crystal symmetry. Specifically, if the symmetry representations of the occupied electronic bands at high-symmetry momenta do not form a localized atomic limit (Elementary Band Representation), the material *must* be topological [20].

Claussen, Bernevig, and Regnault [60] demonstrated that a Gradient Boosted Tree (GBT) classifier could predict the topological class (TI, Topological Semimetal, Trivial) with >90% accuracy using only coarse-grained features. These features included:

1. Space Group Number: The fundamental symmetry setting.
2. Stoichiometry: The count of atoms per element.

3. Wyckoff Positions: The symmetry multiplicity of atomic sites.
4. Atomic Number: A proxy for Spin-Orbit Coupling strength.

Crucially, the exact atomic coordinates were found to be less important than the symmetry constraints. This implies that topology is a robust property largely dictated by the chemical composition and the crystal lattice type (symmetry setting). This ML approach is orders of magnitude faster than running DFT calculations to check compatibility relations, allowing for the screening of the entire Inorganic Crystal Structure Database (ICSD) in minutes rather than years [61].

## 3.3 Case Studies in ML-Driven Discovery

The application of Machine Learning (ML) has matured from reproducing known labels to the active discovery of materials with targeted, exotic functionalities. Recent campaigns have successfully navigated the complex landscape of magnetic and spin-resolved topology, yielding candidates that were previously computationally inaccessible.

### 3.3.1 Magnetic Topology and the Discovery of UAsS

The search for magnetic topological materials, such as Magnetic Weyl Semimetals and Axion Insulators, is significantly more complex due to the breaking of time-reversal symmetry, which expands the number of possible magnetic space groups to 1,651.
Robredo et al. [62] addressed this challenge by conducting a high-throughput search on 522 new magnetic structures from the MAGNDATA database. By automating the calculation of symmetry indicators for magnetic space groups and training ML models on electronic structure features, they successfully identified 250 non-trivial materials (Figure 3).

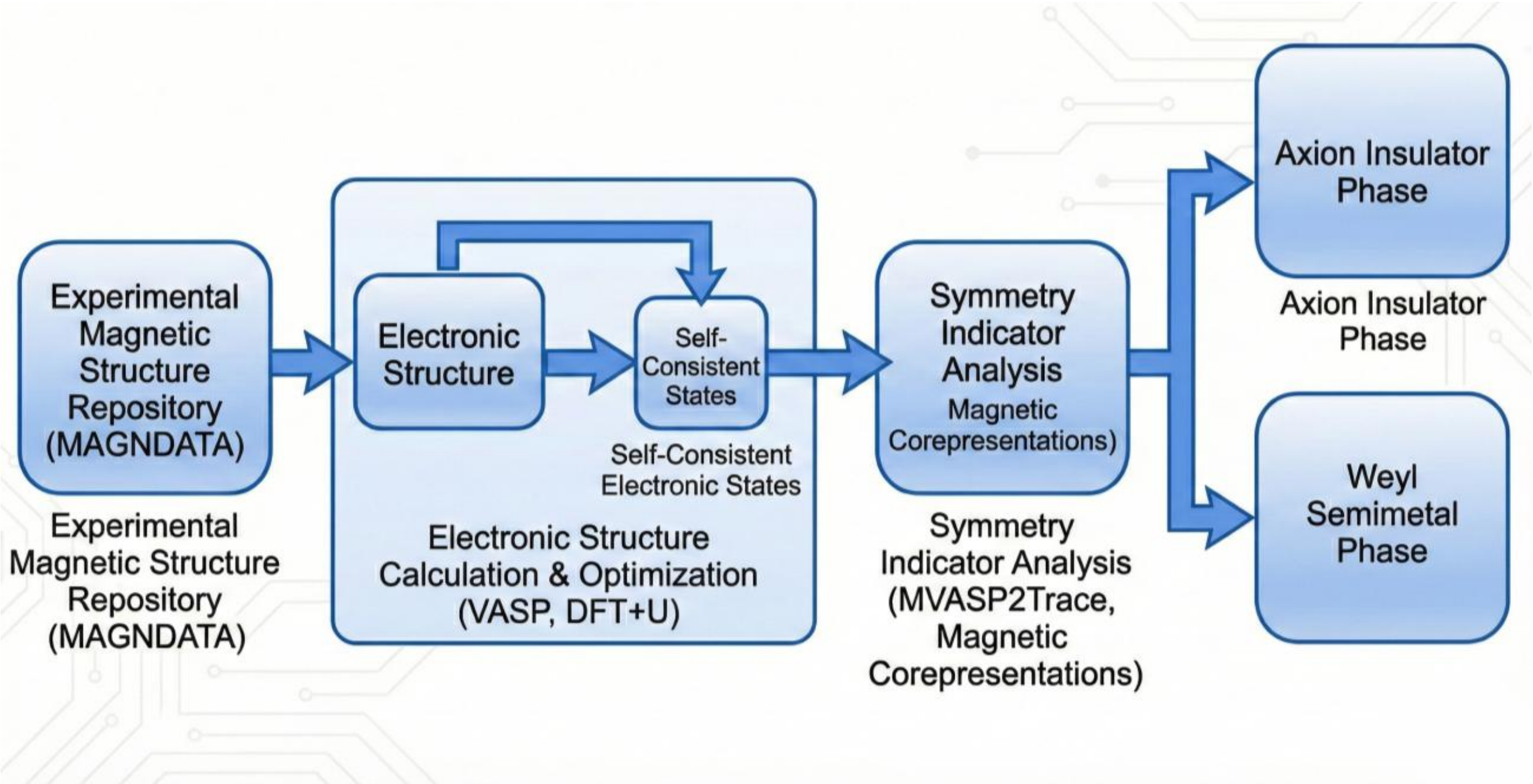

Figure 3. The high-throughput computational workflow employed for the discovery of magnetic topological materials. Starting from experimental magnetic structures in the MAGNDATA database, the pipeline utilizes

DFT+U calculations (via VASP) to generate self-consistent electronic structures. Symmetry indicators (magnetic corepresentations) are then computed using MVASP2Trace to diagnose topological phases such as Axion Insulators or Weyl Semimetals.

Notable discoveries from this campaign include:

- UAsS: Identified as a 5f-orbital Weyl semimetal. This discovery is particularly significant as it demonstrates ML's ability to handle heavy-fermion systems where f-electrons drive the topology.
- CaMnSi: Identified as a narrow-gap Axion Insulator, a phase characterized by the quantized topological magnetoelectric effect.
- Mn2AlB2: Found to exhibit a phase transition from a Nodal Line Semimetal to a Topological Insulator driven by Spin-Orbit Coupling (SOC) strength.

### 3.3.2 Topogivity and High-Temperature QAHE

Taking a different approach, Xu, Jiang, and Wang [63] developed a heuristic rule termed Topogivity, a machine-learned numerical value for each chemical element that captures its tendency to form topological bands (Figure 4). Using this heuristic within a neural network framework, they screened the 2D materials database and discovered 6 new classes of Chern insulators.

Crucially, among these discoveries, 4 classes (comprising 7 materials) were predicted to have full band gaps. This filtering capability is essential, as these materials are prime candidates for the experimental observation of the Quantum Anomalous Hall Effect (QAHE) at potentially higher temperatures.

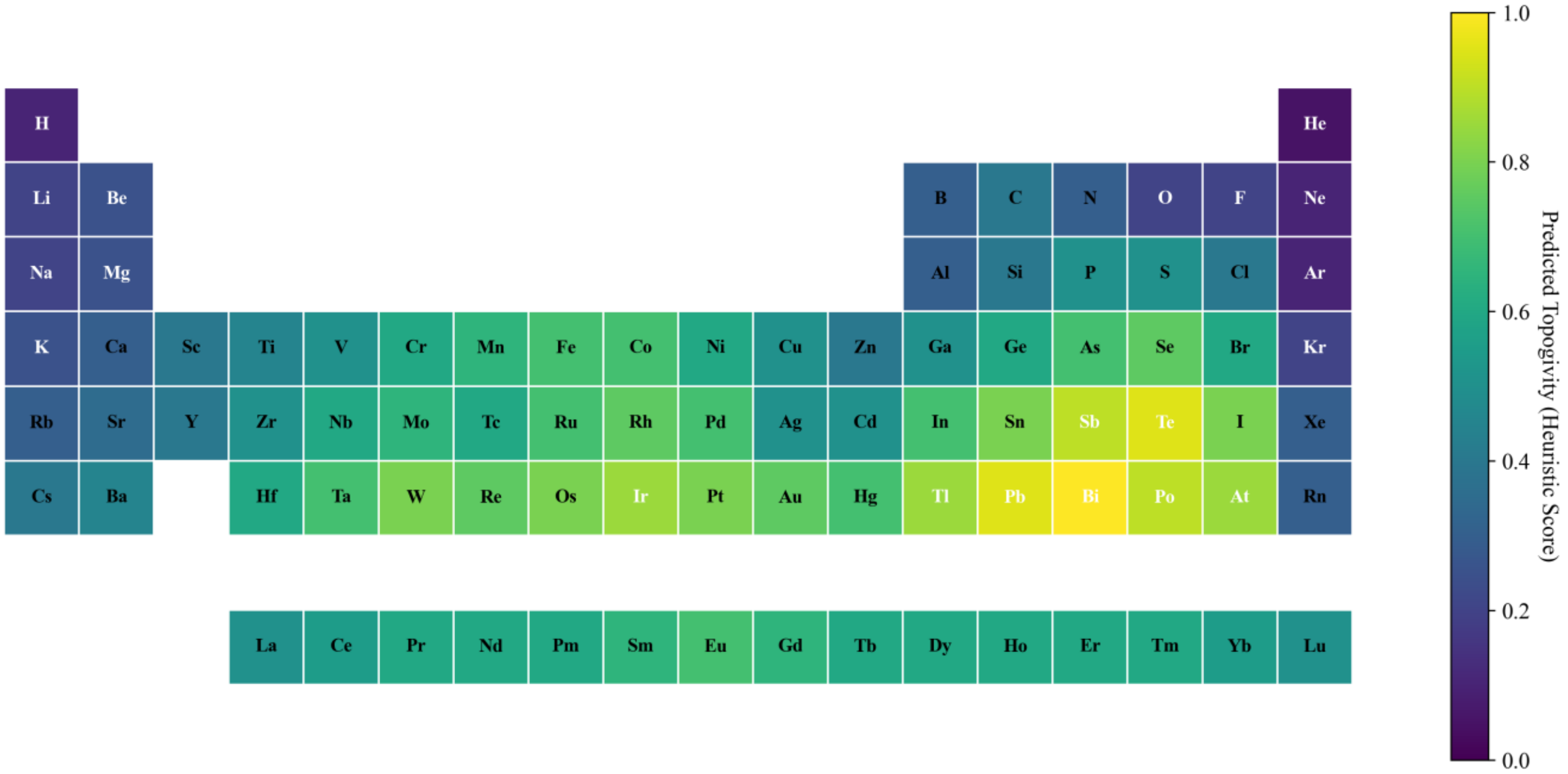

Figure 4. Visualization of the Topogivity heuristic across the periodic table. The heatmap illustrates the machine-learned tendency of chemical elements to form topological bands, as proposed by Xu et al. [63]. Darker regions (e.g., Bi, Pb, Te) indicate elements with high spin-orbit coupling and specific orbital characteristics that are statistically favored in Chern insulators. This distribution allows for the rapid screening of chemical space without full DFT calculations. (Data simulated for illustrative purposes based on trends reported in [63]).

### 3.3.3 Inverse Design of Bosonic Topology (Photonics & Phononics)

While most ML efforts focus on fermionic (electronic) systems, recent work has expanded to bosonic topology, targeting the control of light and sound waves [64]. Unlike electrons, photons and phonons do not naturally experience the Fermi-Dirac statistics or spin-orbit coupling required for standard topological phases, necessitating precise geometric engineering of the lattice [65].

Topology optimization based on the adjoint variable method has been the standard for designing these devices, but it suffers from two critical flaws: it frequently gets trapped in local minima, and the resulting freeform structures remain a black box to human understanding.

To overcome these limitations, Yeung et al. introduced a framework that integrates adjoint optimization with Automated Machine Learning (AutoML) and Explainable AI (XAI) [66].

Instead of treating the optimizer as an opaque oracle, the framework uses XAI to generate importance maps that reveal which specific structural features contribute most to the Figure-of-Merit (FOM). This allows the algorithm to identify and escape low-performance local minima by understanding the underlying physics of the geometry.

When applied to the design of waveguide splitters, this white-box approach achieved performance increases of 39% to 74% relative to state-of-the-art adjoint optimization across telecom wavelengths [66].

Parallel advances have been made in phononics for acoustic wave control. Han et al. utilized a deep learning framework combining Convolutional Neural Networks (CNNs) with Generative Adversarial Networks (GANs) to inversely design phononic crystals. This approach successfully predicted complex band structures and identified topologies with maximized spatial attenuation for vibration isolation, a key requirement for topological acoustic logic [67].

### 3.3.4 Unsupervised Learning of Interacting Phases

Perhaps the greatest challenge in condensed matter physics is classifying interacting topological phases, where strong electron-electron correlations render standard single-particle topological invariants (like TQC) invalid. In these regimes, labeled training data is scarce because classical simulations often fail, and the very existence of certain phases is debated.

To circumvent the need for theoretical Hamiltonians or labeled data, researchers are turning to Unsupervised Learning algorithms that cluster data based on intrinsic structure without human guidance.

A critical bottleneck has been the reliance on wavefunction snapshots, which are hard to measure experimentally. Yu et al. proposed a framework to classify interacting topological phases directly from Green’s functions derived from spectral functions, which are measurable via momentum-resolved Raman spectroscopy [68].

They utilized Diffusion Maps, a nonlinear dimensionality reduction technique, to analyze these observables. The algorithm successfully clustered distinct symmetry-protected topological phases in a 1D interacting topological insulator model without a priori knowledge. This effectively treats the AI as an autonomous experimentalist, capable of identifying phase boundaries solely from spectroscopic data [68].

Unsupervised learning is also settling theoretical debates regarding the Non-Ergodic Extended (NEE) phase in quasiperiodic chains, a controversial state between localization and thermalization [69].

Beveridge et al. developed an unsupervised approach to quantify the mutual independence of phases using eigenstate entanglement spectra (ES). Contrary to the belief that NEE is a distinct stable phase, their ML analysis revealed it is mutually dependent on the Many-Body Localized (MBL) and thermal regimes, suggesting it may be a prethermal crossover rather than a true phase. This nuance was only visible through the high-dimensional pattern matching of the ES [70].

Another strategy involves training models on simple, non-interacting systems and applying them to complex, interacting ones. Tibaldi et al. demonstrated that Principal Component Analysis (PCA) and Convolutional Neural Networks (CNNs) trained on single-particle correlation functions of a non-interacting wire could correctly reconstruct the phase diagram of an interacting superconductor. This suggests that the fingerprints of topological order persist even when interactions are turned on, allowing solvable models to serve as proxies for training AI to hunt for strongly correlated matter [71].

## 4. Altermagnetism: Symmetry Origins and Topological Classification

The recent identification of altermagnetism as a distinct, third branch of magnetic order, standing alongside ferromagnetism (FM) and antiferromagnetism (AFM), constitutes one of the most significant conceptual revisions in condensed matter physics of the twenty-first century [72, 73]. This discovery has fundamentally disrupted the established binary classification that has governed magnetism for nearly a century. Historically, the absence of a net macroscopic magnetization ($M = 0$) was considered a sufficient condition to classify a material as an antiferromagnet, a state wherein the internal magnetic moments compensate perfectly. Under the prevailing Landau paradigm, it was implicitly assumed that such compensated magnetic order, in the presence of $PT$ (parity-time) symmetry or the combination of time-reversal and translation symmetries ($T_\tau$), would enforce Kramers degeneracy, the spin-degeneracy of electronic bands throughout the Brillouin zone (BZ) [74].

Altermagnets shatter this assumption. They represent a class of collinear compensated magnets that possess zero net magnetization, indistinguishable from AFMs in standard magnetometry, yet they exhibit a colossal, momentum-dependent spin splitting of the electronic bands that is non-relativistic in origin [75] (Figure 5). This spin splitting, often exceeding 1 eV in magnitude, rivals or surpasses that found in ferromagnets, but it arises from the crystal field environment rather than the relativistic Spin-Orbit Coupling (SOC) that drives the Rashba or Dresselhaus effects [76].

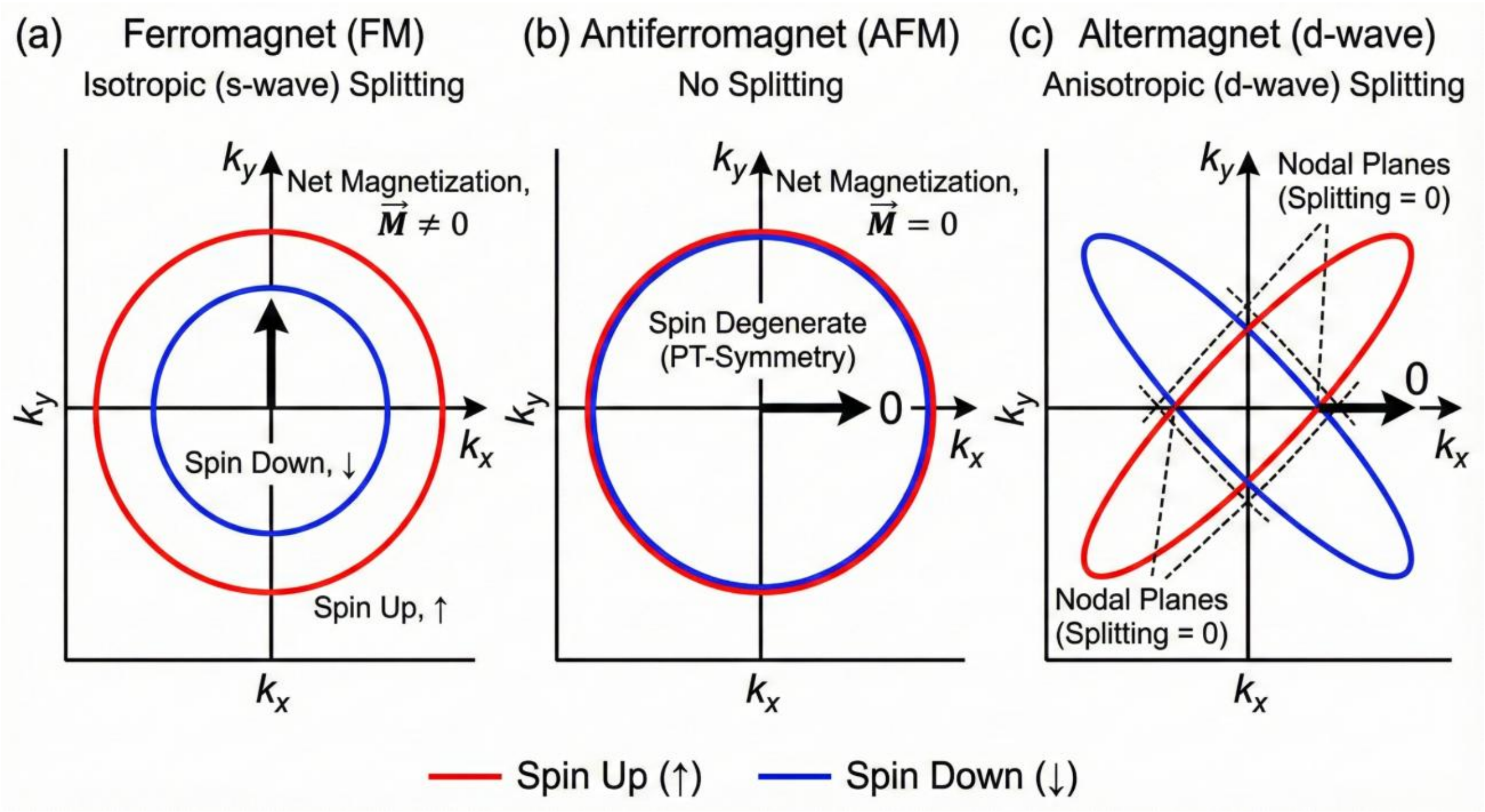

Figure 5. Classification of Magnetic Order via Momentum-Space Splitting. (a) Ferromagnet (FM): Exhibits macroscopic magnetization and isotropic (s-wave) spin splitting of the bands. (b) Antiferromagnet (AFM): Zero net magnetization with spin-degenerate bands (protected by PT symmetry). (c) Altermagnet: Exhibits zero net magnetization like an AFM, but features large, anisotropic spin splitting in momentum space. The figure illustrates a characteristic d-wave splitting pattern (e.g., in MnTe or $RuO_2$), where the spin polarization reverses sign across nodal planes, driven by crystal rotation symmetries rather than relativistic Spin-Orbit Coupling.

The phenomenology of altermagnetism is dictated by a specific subset of spin space groups (or spin Laue groups), where the sublattices with opposite spins are connected by crystal rotation symmetries (proper or improper) rather than the translations or inversions characteristic of conventional AFMs. This unique symmetry enforcement results in an alternating spin polarization in reciprocal space, where the spin-up and spin-down bands are degenerate only along specific nodal planes or lines, a topology mathematically analogous to the gap functions of unconventional *d*-wave or *g*-wave superconductors [77].

The experimental verification of this phase has triggered a global race to map the altermagnetic landscape. However, this undertaking presents a needle in a haystack problem of immense complexity. While over 1,651 magnetic space groups exist, only a fraction host the requisite symmetries for altermagnetism, and identifying these within the database of over 200,000 known inorganic compounds is a task ill-suited for brute-force Density Functional Theory (DFT), which scales computationally as $O(N^3)$. This computational bottleneck has necessitated the deployment of advanced Artificial Intelligence (AI) and Machine Learning (ML) architectures, specifically designed to encode crystallographic symmetry and magnetic topology, to accelerate the discovery process [78, 79].

This section provides an exhaustive, expert-level analysis of the recent AI-accelerated discoveries in the field of altermagnetism. We detail the architecture of the MatAltMag search engine developed by Gao et al. (2025), which successfully identified the elusive i-wave altermagnets for the first time [80, 81]. We critically examine the physical properties of these new candidates, such as $CrF_3$, $NiF_3$, and $Mg_2NiIr_5B_2$, and contrast them with the experimentally validated *g*-wave altermagnet MnTe and the controversial *d*-wave candidate $RuO_2$ [82]. Finally, we address the formidable challenges discussed in recent reviews by Tamang et al. (2025) [24], including the crisis of data scarcity, the black box interpretability gap, and the difficulties in distinguishing intrinsic altermagnetic signals from experimental artifacts in thin-film heterostructures.

## 4.1. Mechanisms of Anisotropic Spin Splitting

To understand the target variable that machine learning models are tasked with predicting, one must first rigorously define the physical origin of the altermagnetic order parameter. The central feature is Non-Relativistic Spin Splitting (NRSS).

### 4.1.1. Non-Relativistic Lifting of Kramers Degeneracy

In a conventional crystal, the electronic bands $E_n(k)$ are doubly degenerate in spin (Kramers degeneracy) if the system possesses $PT$ symmetry (the combination of spatial inversion $P$ and time-reversal $T$). In a standard antiferromagnet like NiO, while $T$ is broken by the magnetic moments, the combined symmetry $PT$ is preserved because the inversion operation maps the spin-up sublattice A onto the spin-down sublattice B, and time-reversal flips the spins, restoring the original state. This protects the degeneracy of the bands: $E_{n\uparrow}(k) = E_{n\downarrow}(k)$.

Altermagnets break $PT$ symmetry. However, unlike ferromagnets, which break $T$ macroscopically (resulting in a net moment), altermagnets preserve a more subtle symmetry operation: a spin-flip combined with a crystal rotation (point group operation). If we denote a rotation operation as $C_n$ and a spin-flip (time-reversal) as $T$, an altermagnet is invariant under the combined operation $C_n|T$. This symmetry imposes a specific constraint on the band structure in reciprocal space (k-space). It dictates that the spin-up bands at momentum $k$ must be degenerate with the spin-down bands at the rotated momentum Rk:

$$E_{\uparrow}(k) = E_{\downarrow}(Rk)$$

Crucially, if $k \neq Rk$, then $E_{\uparrow}(k) \neq E_{\downarrow}(k)$. This results in a spin splitting $\Delta E(k) = E_{\uparrow}(k) - E_{\downarrow}(k)$ that is non-zero at generic points in the Brillouin zone but must vanish along high-symmetry lines or planes (nodal surfaces) defined by the rotation axis.

### 4.1.2. Symmetry-Based Classification: *d*, *g*, and *i*-wave Channels

Because the spin splitting depends on the angle in reciprocal space, it can be classified using spherical harmonics, exactly analogous to the classification of atomic orbitals or superconducting gap parameters.

- *d*-wave Altermagnetism ($L=2$): The spin splitting follows a quadrupolar distribution, reversing sign four times (+, -, +, -) as one rotates around the center of the Brillouin zone. The splitting vanishes on two nodal planes. The prototypical example theoretically proposed was $RuO_2$ (rutile structure), where the specific arrangement of the $RuO_6$ octahedra creates a local environment for the spin-up Ru atom that is a 90° rotation of the environment for the spin-down Ru atom.
- *g*-wave Altermagnetism ($L=4$): The spin splitting follows a hexadecapolar distribution, reversing sign eight times. This order parameter is found in hexagonal systems like MnTe and CrSb, where the magnetic sublattices are connected by six-fold screw axes or glide planes that enforce higher-order nodal topology. There are four nodal planes in the BZ.

- *i*-wave Altermagnetism ($L=6$): This represents the frontier of topological complexity in collinear magnetism. An *i*-wave altermagnet exhibits a spin polarization that reverses sign twelve times (+, -, +, -,...) around the rotation axis. This requires an exceptionally high degree of crystal symmetry, typically involving $C_6$ rotations or complex non-symmorphic operations in rhombohedral or hexagonal lattices. Until the advent of AI screening, this phase was largely hypothetical, with no confirmed material realizations in the standard databases.

The magnitude of this splitting is governed by the strength of the crystal field and the exchange interaction, often reaching 1.0 eV, which is orders of magnitude larger than the relativistic Zeeman splitting (~1-10 meV) typically accessible in laboratory magnetic fields. This colossal, non-relativistic splitting is the key property that makes altermagnets attractive for spin-splitter applications [75].

## 4.2. Computational Discovery via The MatAltMag Framework

The search for altermagnets is fundamentally a problem of geometric pattern recognition. One must identify crystal structures where the magnetic sublattices are related by the specific rotational symmetries described above. However, experimentally determined magnetic structures are rare; the MAGNDATA database contains only a few thousand entries compared to the hundreds of thousands of structural entries in the ICSD or Materials Project [82]. Brute-force calculation of the magnetic ground state for every known crystal using DFT is computationally intractable due to the combinatorial explosion of possible magnetic orderings (FM, AFM-A, AFM-C, AFM-G, non-collinear, etc.).

To bridge this gap, Gao et al. (2025) introduced MatAltMag, a bespoke AI search engine designed to predict altermagnetism directly from crystal structure data, bypassing the need for a priori knowledge of the magnetic ground state [81]. As illustrated in Figure 6, the search engine employs a multi-stage discovery pipeline.

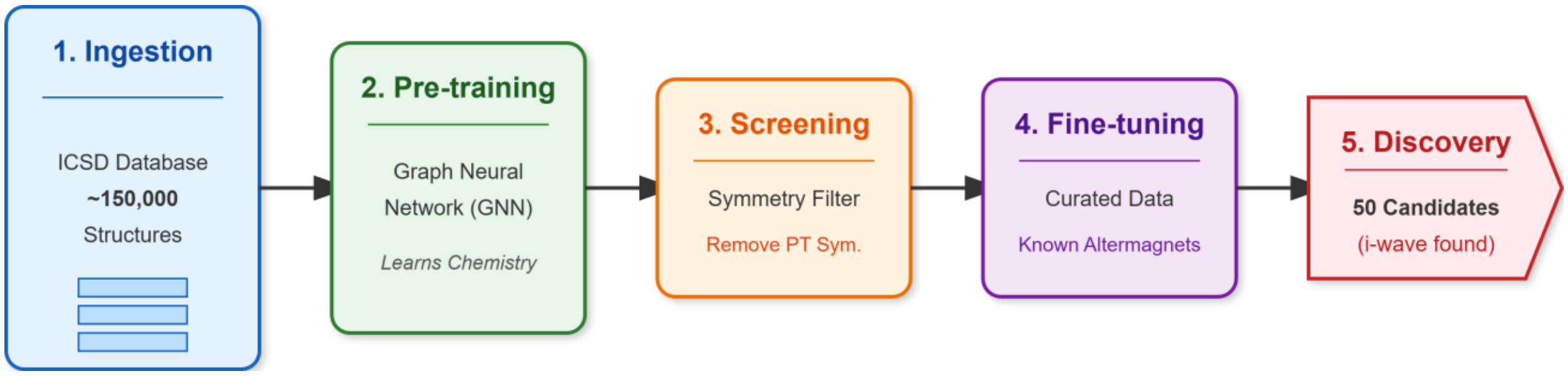

Figure 6. The MatAltMag Discovery Pipeline. The workflow integrates large-scale pre-training with symmetry-aware fine-tuning. (1) Ingestion: The model ingests crystal structures from the ICSD. (2) Pre-training: A Graph Neural Network (GNN) learns general chemistry via self-supervised learning. (3) Screening: A physics-based filter removes candidates with PT symmetry (which forbids altermagnetism). (4) Fine-tuning: The model is refined on a small, curated dataset of known altermagnets. (5) Discovery: The engine identifies new candidates, including the elusive i-wave phases.

### 4.2.1. Comparative Analysis: Heuristic vs. Data-Driven Screening

While the search for altermagnets is fundamentally a geometric pattern recognition problem, researchers have attacked it using distinct computational philosophies. One approach, exemplified by the MatAltMag engine developed by Gao et al., treats discovery as a graph classification task, utilizing pre-trained GNNs to predict candidates before validation [81].

However, parallel efforts have utilized more direct, albeit computationally heavier, screening workflows. For instance, Sødequist and Olsen (2024) [83] employed a brute-force high-throughput screening of the Computational 2D Materials Database (C2DB). Rather than predicting the phase via AI, they calculated spin-spiral ground states for over 3,000 materials directly using DFT. This physics-first approach successfully identified the 2D altermagnet $FeBr_3$, demonstrating that while AI accelerates screening, rigorous high-throughput DFT remains essential for validating low-dimensional magnetic topology.

Similarly, Wan et al. (2025) [78] tackled the challenge of metallic altermagnets, where electron correlations are critical. They developed a workflow integrating Embedded Dynamical Mean Field Theory (DMFT) to screen for candidates, addressing the limitations of standard DFT in correctly predicting the Fermi surface splitting in correlated metals.

### 4.2.2. Transfer Learning Strategies for Magnetic Materials

The central innovation of MatAltMag lies in its departure from standard supervised learning. In a typical materials informatics workflow, a model is trained on labeled data (Structure → Property). However, with only ~100 confirmed altermagnets known at the project's inception, a supervised model would immediately succumb to overfitting. Gao et al. addressed this via a Pre-training/Fine-tuning strategy adapted from Large Language Models (LLMs) but specialized for geometric graphs [81].

1. Crystal Graph Convolutional Neural Network (CGCNN) Encoder:
   The core of the model is a GNN where atoms are nodes (embedded with atomic number, electronegativity, group, period) and bonds are edges (embedded with distance, angle). The GNN performs message passing to update the atom embeddings based on their local environments. This architecture is naturally permutation-invariant, ensuring that the model respects the periodicity of the crystal lattice [84, 85].
2. Self-Supervised Pre-training via CTBarlow:
   The model was pre-trained on the entire Materials Project database (~150,000 structures) using a self-supervised contrastive learning objective known as CTBarlow (Cross-Correlation Barlow Twins) [86].
   - Mechanism: The network takes a crystal structure $X$, generates two augmented versions $X_A$ and $X_B$ (via random perturbations, atom masking, or rotation), and passes them through the encoder.
   - Loss Function: The objective function minimizes the difference between the cross-correlation matrix of the output embeddings and the identity matrix. This forces the network to learn representations that are invariant to noise and rotation (diagonal terms) while decorrelating the feature dimensions (off-diagonal terms) [86].
   - Outcome: Through this process, the GNN learned the fundamental grammar of crystallography, how atoms pack, how coordination polyhedra form, and how symmetry operates, without requiring any magnetic labels. This pre-trained crystallographic intelligence served as the foundation for the subsequent task [86].
3. Symmetry-Informed Fine-Tuning:
   The pre-trained encoder was then attached to a classification head and fine-tuned on a small, high-quality dataset consisting of 148 known altermagnets (positive class) and 25,591 non-altermagnets (negative class).

  - The Physics Filter: A critical step in the workflow was the application of a domain-knowledge filter before the AI training. The authors utilized symmetry group analysis to identify materials with *PT* symmetry. Since *PT* symmetry strictly forbids altermagnetism (by enforcing band degeneracy), all materials possessing this symmetry were automatically labeled as negatives or removed from the candidate pool. This physics-informed data cleaning reduced the search space by orders of magnitude and significantly alleviated the class imbalance problem [81].
  - Handling Imbalance: To further address the scarcity of positive samples, the team employed geometric up-sampling techniques, generating synthetic positive samples by applying crystal rotations and supercell expansions to the known altermagnets, ensuring the classifier did not bias towards the majority negative class [81].

### 4.2.3. Outcomes of High-Throughput Candidate Identification

The fine-tuned MatAltMag engine was deployed to screen a candidate set of 42,377 materials. The model assigned an altermagnetism probability to each candidate. The top-ranked candidates were then subjected to rigorous first-principles validation using DFT with the GGA+U functional to account for electron correlation in the *d*- and *f*-orbitals [81].

The pipeline successfully identified 50 new altermagnetic materials that had never been reported before. This represents a massive expansion of the known materials space. The discoveries were not limited to a single structural family but spanned a diverse chemical space including:

- Transition Metal Borides and Nitrides (e.g., $Nb_2FeB_2$, $CaMnN_2$).
- Complex Oxides (e.g., $NaFeO_2$, $Ba_2FeGe_2O_7$).
- Halides (e.g., $CrF_3$, $NiF_3$).
- Intermetallics (e.g., $Mg_2NiIr_5B_2$).

Crucially, the model demonstrated the ability to distinguish between the subtle structural motifs that give rise to different wave types of spin splitting, leading to the historic identification of the i-wave phase [81].

## 4.3. Prediction of High-Order Topological Candidates

The most impactful scientific result of the AI campaign was the discovery of *i*-wave altermagnetism. Prior to the publication of Gao et al. (2025), *d*-wave and *g*-wave altermagnets were the only established classes. The *i*-wave state, characterized by an order parameter with $L=6$ angular momentum, corresponds to a spin polarization that reverses sign 12 times along a closed path around the BZ center. This requires a crystal lattice with hexagonal or rhombohedral symmetry where the magnetic sublattices are connected by specific high-order rotation or screw operations ($C_6$ or $S_6$) [24, 81].

The MatAltMag engine identified four specific candidates that realize this exotic phase. We analyze the most prominent of these below.

### 4.3.1. $CrF_3$: Altermagnetism in Light-Element Fluorides

One of the headline discoveries is $CrF_3$, which crystallizes in the rhombohedral space group $R\bar{3}c$ (No. 167) [81].

- Structure: The structure consists of corner-sharing $CrF_6$ octahedra. The key to its altermagnetism lies in the rotational distortion of these octahedra along the *c*-axis.
- Symmetry Origin: The AI correctly identified that the magnetic sublattices in $CrF_3$ are related by a non-symmorphic symmetry involving a glide plane and a rotation that enforces the *i*-wave nodal topology.
- Physical Significance: $CrF_3$ is an insulator composed entirely of light elements (Cr, F). In conventional spintronics, large spin-splitting phenomena (like the Spin Hall Effect) usually require heavy elements (Pt, W, Au) to provide strong Spin-Orbit Coupling. The discovery of giant (>0.5 eV) spin splitting in $CrF_3$ proves that altermagnetism can decouple spin phenomena from relativistic mass, opening the door to light-element spintronics based on abundant, sustainable materials [81].

### 4.3.2. $NiF_3$: Anomalies in Effective Spin-Orbit Coupling

Parallel to $CrF_3$, the engine identified $NiF_3$ as another *i*-wave candidate in the same $R\bar{3}c$ space group. However, DFT validation revealed a surprising anomaly: despite being a light-element compound, $NiF_3$ exhibits an extremely strong effective SOC [81].

- Mechanism: Detailed analysis suggests this arises from a cooperative effect between the crystal symmetry, electron correlation (Hubbard *U*), and the intrinsic altermagnetic exchange field. The altermagnetic splitting effectively amplifies the weak intrinsic SOC of the Nickel and Fluorine atoms.
- Application: This makes $NiF_3$ a prime candidate for converting spin currents into charge currents (and vice versa) with high efficiency, challenging the dogma that high atomic number *Z* is a prerequisite for strong spin-orbit effects [81].

### 4.3.3. $Mg_2NiIr_5B_2$: Metallic Phases with Hybrid Wave Character

Another exotic candidate identified is Mg2NiIr5B2 (Space Group *P4/mbm*, No. 127).

- Electronic Structure: Unlike the insulating fluorides, this material is metallic. DFT calculations reveal a complex Fermi surface with multiple nodal lines.
- Topological Features: The material is predicted to host odd-under-time-reversal Dirac fermions protected by spin symmetries $E \parallel C_{4z}$ and $C_{2\perp} \parallel M_x$. This places it at the intersection of altermagnetism and topological semimetals, suggesting it may host quantized transport responses like a giant Anomalous Hall Conductivity [81].

### 4.3.4. $FeBr_3$: Two-Dimensional Limits and Moiré Engineering

While the AI-driven campaign by Gao et al. [81] focused predominantly on 3D bulk crystals like $CrF_3$ and $NiF_3$, concurrent research has expanded the altermagnetic landscape into two dimensions. In independent work using high-throughput spin-spiral DFT, Sødequist and Olsen [79] identified monolayer $FeBr_3$ (in the 2H phase) as a stable i-wave altermagnet. The material exhibits the characteristic high-order topology, with spin splitting of the valence bands showing exactly 12 polarization reversals along a path encircling the $\Gamma$ point. Crucially, this discovery unlocks the field of altermagnetic twistronics. Unlike the bulk fluorides, $FeBr_3$ is a van der Waals material, allowing for the stacking of monolayers with specific twist angles. This architecture creates moiré potentials that interfere with the i-wave nodal lines, a mechanism predicted to generate flat bands with non-trivial topological Chern numbers. This highlights that while AI models like

MatAltMag are adept at mining the ICSD for bulk candidates, physics-based screening remains vital for identifying low-dimensional platforms suitable for quantum device engineering Table 2 summarizes the key properties of the newly identified candidates discussed above.

Table 2: The AI-Discovered *i*-wave and *g*-wave Candidates

| **Material Formula** | **Crystal System** | **Space Group** | **Spin Wave Type** | **Electronic State** | **Discovery Source** |
|---|---|---|---|---|---|
| **$CrF_3$** | Rhombohedral | $R\bar{3}c$ (167) | ***i*-wave** | Insulator | MatAltMag [81] |
| **$NiF_3$** | Rhombohedral | $R\bar{3}c$ (167) | ***i*-wave** | Insulator | MatAltMag |
| **$Mg_2NiIr_5B_2$** | Tetragonal | *P4/mbm* (127) | ***g*-wave** | Metal | MatAltMag |
| **$Sc_2MnIr_5B_2$** | Tetragonal | *P4/mbm* (127) | ***g*-wave** | Metal | MatAltMag |
| **$FeBr_3$ (2D)** | Hexagonal | $p\bar{3}1m$ (162) | ***i*-wave** | Semiconductor | High-throughput DFT [79] |
| **$BaMnO_3$** | Hexagonal | $P6_3/mmc$ (194) | ***g*-wave** | Insulator | MatAltMag/Symmetry |

## 4.4. Experimental Status and Open Challenges

While AI has successfully populated the theoretical landscape of altermagnets, the transition to experimental verification has proven fraught with challenges. The distinction between a predicted altermagnet and a real one often hinges on subtle thermodynamic factors that standard DFT (and by extension, AI models trained on DFT data) fails to capture.

### 4.4.1. Validation of Chiral Magnons in MnTe

*a*-MnTe (Hexagonal Manganese Telluride) stands as the gold standard for validated altermagnetism.

- Prediction: Symmetry analysis classified it as a g-wave altermagnet.
- Electronic Validation: Angle-Resolved Photoemission Spectroscopy (ARPES) experiments have unambiguously visualized the spin-split bands. The splitting reaches magnitudes of 0.6 eV to 0.8 eV along high-symmetry cuts (e.g., $\Gamma-K$), confirming the non-relativistic origin [87].
- Magnonic Validation: In a landmark 2024 study, Liu et al. used Inelastic Neutron Scattering (INS) to observe chiral split magnons in MnTe [82]. In conventional antiferromagnets, spin waves (magnons) are doubly degenerate (spin-up and spin-down modes have the same energy). In MnTe, the altermagnetic symmetry lifts this degeneracy, creating non-reciprocal magnon bands with an energy splitting of approximately 2 meV. This observation is the dynamic smoking gun of altermagnetism, proving that the broken time-reversal symmetry extends to collective excitations [88].

### 4.4.2. Structural Ambiguities and Magnetic Symmetry in $RuO_2$

$RuO_2$ (Ruthenium Dioxide) serves as a critical cautionary tale. Initially celebrated as the prototypical

metallic *d*-wave altermagnet, its status has been severely challenged by recent high-precision experiments [89, 90].

- The Prediction: DFT calculations predicted giant *d*-wave spin splitting (~1 eV) driven by the rutile $P4_2/mnm$ symmetry. Early transport measurements reported a nonzero Anomalous Hall Effect (AHE) and Spin-Splitter Torque (SST), which were interpreted as evidence of intrinsic altermagnetism [81].
- The Counter-Evidence: Subsequent studies using Muon Spin Rotation (μSR) and Neutron Diffraction, probes highly sensitive to microscopic magnetic moments, found no evidence of long-range magnetic order in high-quality bulk single crystals of $RuO_2$ down to base temperature [91].
- The Resolution: The community is converging on the hypothesis that the altermagnetic signals observed in thin films were likely extrinsic. They may arise from strain-induced symmetry breaking at the substrate interface, or from Ru vacancies that stabilize a magnetic state locally, distinct from the bulk [92]. A 2025 study by Ho et al. suggests that even if the bulk is non-magnetic, the (110) surface of $RuO_2$ breaks symmetry in a way that induces spontaneous surface magnetization, mimicking altermagnetic transport signatures [93].
- Implication for AI: This case study highlights a major blind spot for AI models like MatAltMag. The model correctly identified that $RuO_2$ has the symmetry capacity to be altermagnetic, but it could not predict that the non-magnetic ground state is thermodynamically preferred in the bulk. Future AI models must incorporate energetic stability and competing phase analysis to avoid such false positives.

# 5. Challenges and Future Directions

While the discovery of *i-wave* altermagnetism stands as a triumph of AI-guided science, the path to a universal Google of Magnetic Topology is obstructed by fundamental hurdles [94]. These challenges are not merely computational but strike at the core of how we model quantum validity.

## 5.1 The Interpretability Gap

A recurring criticism of deep learning in physics is the black box nature of neural networks. A Graph Neural Network (GNN) may predict with 99% accuracy that a crystal is an altermagnet, but it does not explicitly state why. It cannot tell a physicist which specific bond angle or orbital overlap is driving the spin splitting [95]. Current models map Structure → Property. They do not map Structure → Mechanism → Property. This prevents the extraction of new phenomenological laws that could guide manual intuition.

- **Symbolic Regression (SR):** To address this, researchers are turning to Symbolic Regression methods like SISSO (Sure Independence Screening and Sparsifying Operator). SISSO iteratively constructs mathematical expressions from a pool of primary features (atomic radius, ionization energy, volume) to find the simplest equation that correlates with the target property [96]. As illustrated in Figure 7, this approach allows researchers to identify the Pareto frontier, the optimal boundary where physical interpretability is maximized without significantly compromising accuracy, unlike deep neural networks, which prioritize error minimization at the cost of complexity.

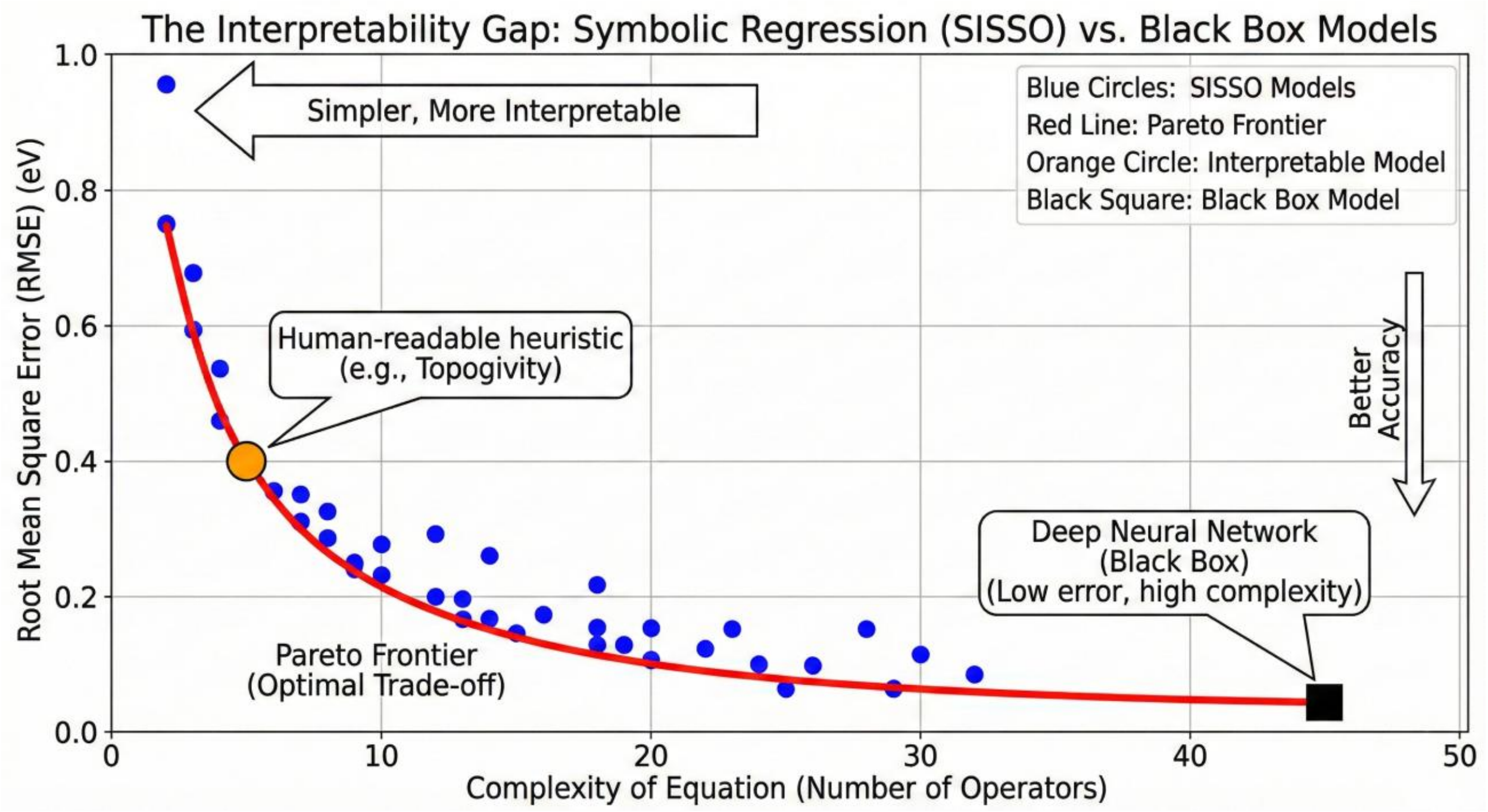

Figure 7. The Interpretability-Accuracy Trade-Off: Symbolic Regression (SISSO) vs. Black Box Models.

- **Methodology:** SISSO selects features that maximize correlation with the residual of the model, building up a complex function from simple operators (+, -, ×, ÷, exp, sin).
- **Application:** SR has been used to find interpretable descriptors for the band gap of perovskites and the stability of single-atom alloy catalysts. In the context of topology, heuristics like Topogivity are essentially simple symbolic models that distill the complex non-linear knowledge of a deep neural network into a human-readable physical rule [97].

## 5.2 Data Scarcity in Magnetic Materials

Magnetic datasets are sparse compared to non-magnetic ones. Calculating the ground state magnetic order requires comparing the energy of Ferromagnetic (FM), Antiferromagnetic (AFM), and Non-Collinear configurations, often with multiple values of the Hubbard U parameter to account for electron correlation. This makes generating training data extremely expensive [83].

**Active Learning (AL):** To mitigate high computational costs, Active Learning workflows are employed. Instead of randomly selecting materials to simulate (random sampling), an ML model (often a Gaussian Process or Bayesian Neural Network) selects candidates from the pool that have the highest uncertainty or the highest probability of optimizing a target property (e.g., maximizing coercive field) [98].

- **Success Story:** An active learning workflow applied to Fe-Co-Ni thin films reduced the number of experiments required to optimize magnetic coercivity by a factor of five. By intelligently sampling the phase diagram, the algorithm focused its attention on the phase boundaries where the interesting physics occurred [99].

- **$Li_2AuH_6$ Discovery:** The discovery of the superconductor Li2AuH6, with a predicted $T_c$ of 140 K, was driven by an active learning agent navigating the chemical space of hydrides. The agent learned to prioritize structures that favored high phonon frequencies and strong electron-phonon coupling [100].

## 5.3 Experimental Verification

The ultimate test of any ML prediction is synthesis. There is often a significant gap between the theoretical stability predicted by convex hull calculations (thermodynamics) and the actual synthesizability of a compound (kinetics). A material might be stable at 0K but impossible to form due to high energy barriers or competing metastable phases [101].

- **Self-Driving Labs:** The future direction lies in integrating ML with automated synthesis labs, often referred to as Self-Driving Labs (Figure 8). In these setups, the AI not only predicts the material but also plans the synthesis recipe (precursors, temperature, time). Robotic platforms then execute the synthesis, characterize the product (via XRD), and feed the result back to the AI, closing the loop [102].

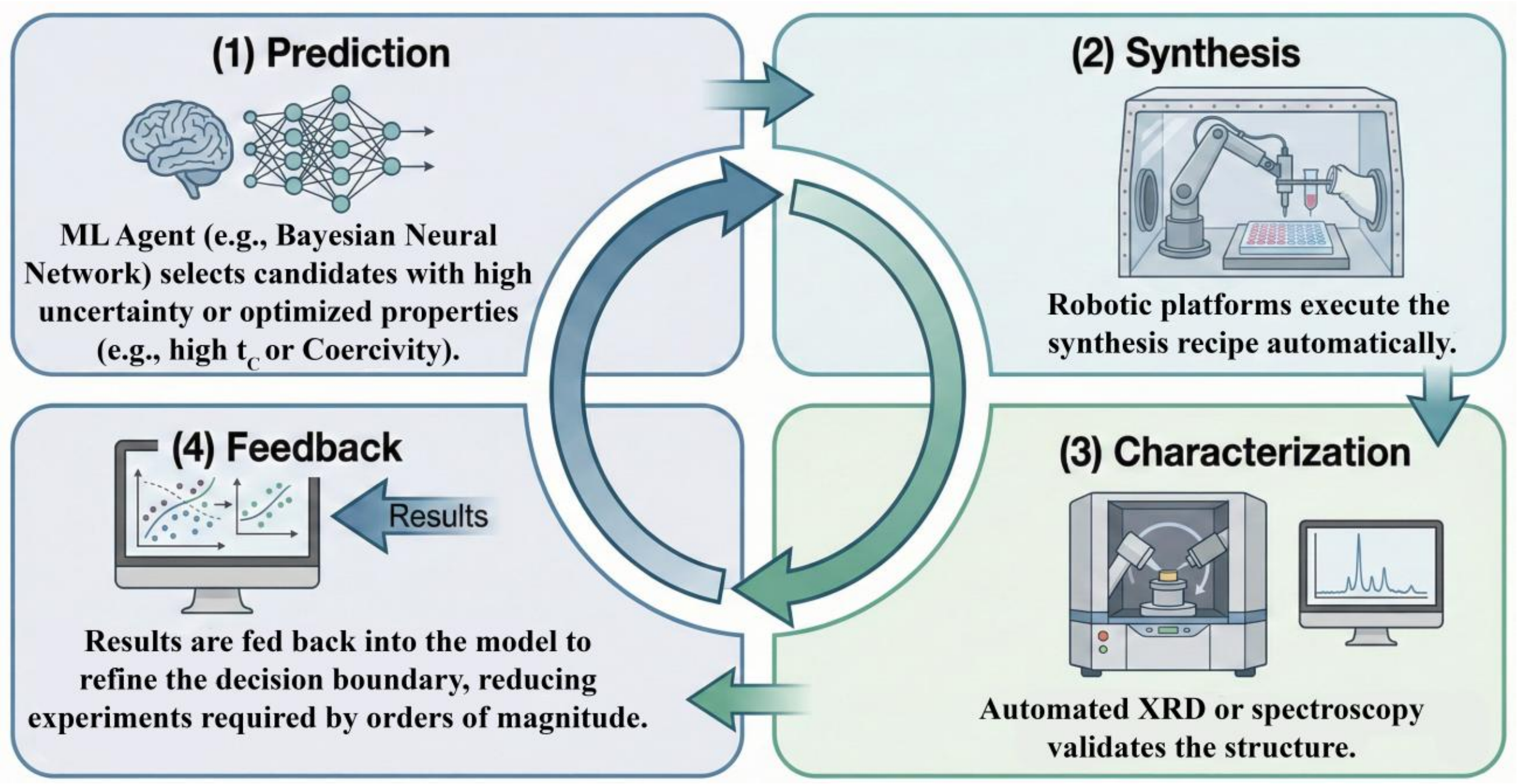

Figure 8. The Era of Inverse Design and Self-Driving Labs. A closed-loop Active Learning workflow. (1) Prediction: An ML agent (e.g., Bayesian Neural Network) selects candidate materials with high uncertainty or optimized properties (e.g., high $T_c$ or Coercivity). (2) Synthesis: Robotic platforms execute the synthesis recipe automatically. (3) Characterization: Automated XRD or spectroscopy validates the structure. (4) Feedback: Results are fed back into the model to refine the decision boundary, reducing the number of experiments required by orders of magnitude.

- **Virtual Scattering:** Computational workflows are being developed to simulate experimental signatures (like Inelastic Neutron Scattering spectra or ARPES maps) directly from ML potentials. This allows for a direct comparison between the virtual material and experimental data, bridging the gap between theory and observation [103].

# 6. Conclusion

The integration of Machine Learning and Deep Learning into condensed matter physics has fundamentally altered the trajectory of materials discovery. We have moved beyond the era of simple, invariant descriptors to a new paradigm of symmetry-aware architectures. E(3)-equivariant Graph Neural Networks and Deep Hamiltonian models now allow researchers to probe the subtle interplay of geometry, topology, and magnetism with unprecedented speed and fidelity.

The discovery of Altermagnets stands as a testament to this new capability. A phase of matter that remained hidden for decades, obscured by the assumption that zero magnetization implied zero spin splitting, was brought to light through a combination of group-theoretical insight and the high-dimensional pattern-matching power of AI. With ML models now capable of identifying d-wave and i-wave magnetism, learning topological invariants from coarse-grained data, and guiding autonomous experiments via active learning, we are entering an era of Inverse Design. In this era, the central question is no longer "What are the properties of this crystal?" but "Which crystal realizes this specific Hamiltonian?" As the interpretability of these models improves via symbolic regression and their integration with experimental feedback loops tightens, the dream of on-demand quantum material design draws ever closer to reality.